\documentclass[aps,showpacs,preprintnumbers,nofootinbib, epsf]{revtex4}
\usepackage{graphicx}

\begin{document}

\preprint{gr-qc/0505034}
\title{Chaplygin gas dominated anisotropic brane world cosmological models}
\author{M. K. Mak}
\email{mkmak@vtc.edu.hk} \affiliation{Department of Physics, The
University of Hong Kong, Pok Fu Lam Road, Hong Kong}
\author{T. Harko}
\email{harko@hkucc.hku.hk} \affiliation{Department of Physics, The
University of Hong Kong, Pok Fu Lam Road, Hong Kong}
\date{\today}

\begin{abstract}
We present exact solutions of the gravitational field equations in the
generalized Randall-Sundrum model for an anisotropic brane with Bianchi type
I geometry, with a generalized Chaplygin gas as matter source. The
generalized Chaplygin gas, which interpolates between a high density
relativistic era and a non-relativistic matter phase, is a popular dark
energy candidate. For a Bianchi type I space-time brane filled with a
cosmological fluid obeying the generalized Chaplygin equation of state the
general solution of the gravitational field equations can be expressed in an
exact parametric form, with the comoving volume taken as parameter. In the
limiting cases of a stiff cosmological fluid, with pressure equal to the
energy density, and for a pressureless fluid, the solution of the field
equations can be expressed in an exact analytical form. The evolution of the
scalar field associated to the Chaplygin fluid is also considered and the
corresponding potential is obtained. The behavior of the observationally
important parameters like shear, anisotropy and deceleration parameter is
considered in detail.
\end{abstract}

\maketitle



\section{Introduction}

The observations of high redshift supernovae \cite{Pe99} and the
Boomerang/Maxima data \cite{Ber00}, showing that the location of the first
acoustic peak in the power spectrum of the microwave background radiation is
consistent with the inflationary prediction $\Omega =1$, have provided
compelling evidence for a net equation of state of the cosmic fluid lying in
the range $-1\leq w=p/\rho <-1/3$. To explain these observations, two dark
components are invoked: the pressureless cold dark matter (CDM) and the dark
energy (DE) with negative pressure. CDM contributes $\Omega _{m}\sim 0.3$%
,and is mainly motivated by the theoretical interpretation of the galactic
rotation curves and large scale structure formation. DE is assumed to
provide $\Omega _{DE}\sim 0.7$ and is responsible for the acceleration of
the distant type Ia supernovae. There are a huge number of candidates for DE
in the literature (for recent reviews see \cite{PeRa03} and \cite{Pa03}).
However, neither CDM nor DE has direct laboratory observational or
experimental evidence for its existence. It would be important if a unified
dark matter -dark energy scenario could be found, in which these two
components are different manifestations of a single fluid \cite{Paetal}.

A candidate for such an unification is the so-called generalized
Chaplygin gas, which is an exotic fluid with the equation of state
$p=-B/\rho ^{\alpha }$, where $B$ and $\alpha $ are two parameters
to be determined. It was initially suggested by Kamenshchik,
Moschella and Pasquier \cite{Ka01} with $\alpha =1$, and then
generalized by Bento, Bertolami and Sen \cite{Be02} for the case
$\alpha \neq 1$. The Chaplygin gas also appears in the
stabilization of branes in
Schwarzschild-AdS black hole bulks as a critical theory at the horizon \cite%
{Ka00} and in the stringy analysis of black holes in three dimensions \cite%
{Ka98}. The Chaplygin equation of state can also be derived from Born-Infeld
type Lagrangians \cite{Be02}, \cite{No05}. This simple and elegant model
smoothly interpolates between a non-relativistic matter phase ($p=0$) and a
negative-pressure dark energy dominated phase.

The cosmological implications of the Chaplygin gas model have been
intensively investigated in the recent literature \cite{Chco}.

The Chaplygin gas cosmological model has been constrained by using
different cosmological observations, like type Ia supernovae
\cite{Ber04}, \cite{Fa02}, the CMB
anisotropy measurements \cite{Be03}, gravitational lensing surveys \cite%
{De03}, the age measurement of high redshift objects \cite{Al03} and the
X-ray gas mass fraction of clusters \cite{Cu04}. The obtained results are
somewhat controversial, with some of them claiming good agreement between
the data and the Chaplygin gas model, while the rest ruling it as a feasible
candidate for dark matter. In particular, the standard Chaplygin gas model
with $\alpha =1$ is ruled out by the data at a 99\% level \cite{Cu04}.

The embedding approach to gravity, according to which our
four-dimensional space-time is a three-brane, embedded in a
five-dimensional space-time (the bulk), has a long history,
beginning at least with the classic book by Eisenhart \cite{Ei}.
More recently, this idea was considered by Akama \cite{Ak83} and
Rubakov and Shaposhnikov \cite{RuSh83} (an exhaustive list of
references on the early articles published in this field can be
found in Bandos and Kummer \cite{BaKu99} and Pavsic and Tapia
\cite{PaTa00}, respectively). Recently, due to the proposal by
Randall and Sundrum \cite{RS99a} the idea of the embedding of our
Universe in a higher dimensional space had attracted again a
considerable interest. According to the brane-world scenario, the
physical fields (electromagnetic, Yang-Mills etc.) in our
four-dimensional Universe are confined to the three brane. These
fields are assumed to arise as fluctuations of branes in string
theories. Only gravity can freely propagate in both the brane and
bulk space-times, with the gravitational self-couplings not
significantly modified. This model originated from the study of a
single $3$-brane
embedded in five dimensions, with the $5D$ metric given by $%
ds^{2}=e^{-f(y)}\eta _{\mu \nu }dx^{\mu }dx^{\nu }+dy^{2}$, which, due to
the appearance of the warp factor, could produce a large hierarchy between
the scale of particle physics and gravity. Even if the fifth dimension is
uncompactified, standard $4D$ gravity is reproduced on the brane. Hence this
model allows the presence of large, or even infinite non-compact extra
dimensions. Our brane is identified to a domain wall in a $5$-dimensional
anti-de Sitter space-time. For a review of dynamics and geometry of brane
Universes see \cite{Ma01}.

Due to the correction terms coming from the extra dimensions, significant
deviations from the Einstein theory occur in brane world models at very high
energies \cite{SMS00}. Gravity is largely modified at the electro-weak scale
$1$ TeV. The cosmological implications of the brane world theories have been
extensively investigated in the physical literature \cite{all}.
Gravitational collapse can also produce high energies, with the five
dimensional effects playing an important role in the formation of black
holes \cite{all1}.

The possibility that the non-local effects, generated by the free
gravitational field of the bulk in a conformally symmetric brane, may
provide an explanation for the dynamics of the neutral hydrogen clouds at
large distances from the galactic center, and hence provide a purely
gravitational explanation for the existence of dark matter, was suggested in %
\cite{MaHa04}.

The behavior and the properties of the anisotropic Bianchi type I
brane-world for different type of matter sources, including inflationary
scalar fields, have been intensively investigated in the framework of the
brane world cosmologies \cite{MSS00}. The study of anisotropic homogeneous
brane world cosmological models have shown an important difference between
these models and standard four-dimensional general relativity, namely, that
brane Universes are born in an isotropic state. For Bianchi type I and V
geometries this type of behavior has been found both by exactly solving the
gravitational field equations \cite{ChHaMa01a}, or from the qualitative
analysis of the model \cite{CS201}. A general analysis of the anisotropy in
spatially homogeneous brane world cosmological models has been performed by
Coley \cite{Co01a}, who has shown that the initial singularity is isotropic,
and hence the initial conditions problem is naturally solved. Consequently,
close to the initial singularity, these models do not exhibit Mixmaster or
chaotic-like behavior \cite{Co01b}. Based on the results of the study of
homogeneous anisotropic cosmological models Coley conjectured that the
isotropic singularity could be a general feature of brane cosmology. The
analytic study of the values of the anisotropy in the limit of small times
shows that for all Bianchi type space-times filled with a non-zero pressure
cosmological fluid, obeying a barotropic equation of state, the initial
singular state on the brane is isotropic \cite{HaMa04a}. This result is
obtained by assuming that in the limit of small times the asymptotic
behavior of the scale factors is of Kasner-type. For brane worlds filled
with dust or with a scalar field, the initial value of the anisotropy
coincides in both brane world and standard four-dimensional general
relativistic cosmologies.

It is the purpose of this paper to investigate some classes of exact
solutions of the gravitational field equations in the brane world model for
the anisotropic Bianchi type I geometry, by assuming that the matter source
on the brane consist of a generalized Chaplygin gas. In this case the
general solution of the field equations can be obtained in an exact
parametric form, for arbitrary values of the parameters characterizing the
equation of state of the matter. The inclusion of the quadratic terms in the
energy-momentum tensor of the perfect cosmological fluids leads to major
changes in the early dynamics of the anisotropic Universe, as compared to
the standard general relativistic case. We also consider the brane dynamics
of the scalar field associated to the Chaplygin case and obtain the
corresponding self-interaction potential. The behavior of the
observationally important physical quantities like anisotropy, shear and
deceleration parameters is considered in detail for all these models.

The present paper is organized as follows. The gravitational field equations
for a Chaplygin gas on a brane with an anisotropic Bianchi type I geometry
are written down in Section II. In Section III we present the general
solution of the field equations. In Section III we discuss and conclude our
results.

\section{Geometry, matter and field equations on the brane}

In the $5D$ space-time the brane-world is located as $Y(X^{I})=0$, where $%
X^{I},\,I=0,1,2,3,4$ are $5$-dimensional coordinates. The effective action
in five dimensions is \cite{MW00}
\begin{equation}
S=\int d^{5}X\sqrt{-g_{5}}\left( \frac{1}{2k_{5}^{2}}R_{5}-\Lambda
_{5}\right) +\int_{Y=0}d^{4}x\sqrt{-g}\left( \frac{1}{k_{5}^{2}}K^{\pm
}-\lambda +L^{\text{matter}}\right) ,
\end{equation}
where $k_{5}^{2}=8\pi G_{5}$ is the $5$-dimensional gravitational coupling
constant, $\Lambda _5$ is the vacuum energy in the bulk, $\lambda $ is the
vacuum energy on the brane and where $x^{\mu },\,\mu =0,1,2,3$ are the
induced $4$-dimensional brane world coordinates. $R_{5}$ is the $5D$
intrinsic curvature in the bulk and $K^{\pm }$ is the intrinsic curvature on
either side of the brane.

On the $5$-dimensional space-time (the bulk), with the negative vacuum
energy $\Lambda _5$ and brane energy-momentum as source of the gravitational
field, the Einstein field equations are given by
\begin{equation}
G_{IJ} = k_5^2 T_{IJ}, \qquad T_{IJ} = -\Lambda_5 g_{IJ} + \delta(Y) \left[
-\lambda g_{IJ} + T_{IJ}^{\text{matter}} \right],
\end{equation}
In this space-time a brane is a fixed point of the $Z_2$ symmetry. In the
following capital Latin indices run in the range $0,...,4$ while Greek
indices take the values $0,...,3$.

Assuming a metric of the form $ds^{2}=(n_{I}n_{J}+g_{IJ})dx^{I}dx^{J}$, with
$n_{I}dx^{I}=d\chi $ the unit normal to the $\chi =\text{const.}$
hypersurfaces and $g_{IJ}$ the induced metric on $\chi =\text{const.}$
hypersurfaces, the effective four-dimensional gravitational equations on the
brane (which are the consequence of the Gauss-Codazzi equations) take the
form \cite{SMS00}:
\begin{equation}  \label{Ein}
G_{\mu \nu }=-\Lambda g_{\mu \nu }+k_{4}^{2}T_{\mu \nu }+k_{5}^{4}S_{\mu \nu
}-E_{\mu \nu },
\end{equation}
where $S_{\mu \nu }$ is the local quadratic energy-momentum correction
\begin{equation}
S_{\mu \nu }=\frac{1}{12}TT_{\mu \nu }-\frac{1}{4}T_{\mu }{}^{\alpha }T_{\nu
\alpha }+\frac{1}{24}g_{\mu \nu }\left( 3T^{\alpha \beta }T_{\alpha \beta
}-T^{2}\right) ,
\end{equation}
and $E_{\mu \nu }$ is the nonlocal effect from the bulk free gravitational
filed, transmitted projection of the bulk Weyl tensor $C_{IAJB}$
\begin{equation}
E_{IJ}=C_{IAJB}n^{A}n^{B},\qquad E_{IJ}\rightarrow E_{\mu \nu }\delta
_{I}^{\mu }\delta _{J}^{\nu }\quad \text{as}\quad \chi \rightarrow 0.
\end{equation}

The four-dimensional cosmological constant, $\Lambda $, and the coupling
constant, $k_{4}$, are given by
\begin{equation}
\Lambda =\frac{k_{5}^{2}}{2}\left( \Lambda _{5}+\frac{k_{5}^{2}\lambda ^{2}}{%
6}\right) ,\qquad k_{4}^{2}=\frac{k_{5}^{4}\lambda }{6}.
\end{equation}

The Einstein equation in the bulk, Codazzi equation, also implies the
conservation of the energy momentum tensor of the matter on the brane,
\begin{equation}
D_{\nu }T_{\mu }{}^{\nu }=0.  \label{DT}
\end{equation}

Moreover, the contracted Bianchi identities on the brane imply that the
projected Weyl tensor should obey the constraint
\begin{equation}
D_{\nu }E_{\mu }{}^{\nu }=k_{5}^{4}D_{\nu }S_{\mu }{}^{\nu }.  \label{DE}
\end{equation}

Finally, the equations (\ref{Ein}), (\ref{DT}) and (\ref{DE}) give the
complete set field equations for the brane gravitational field.

For any matter fields (scalar field, perfect fluids, kinetic
gases, dissipative fluids etc.) the general form of the brane
energy-momentum tensor can be covariantly given as \cite{Ma01}
\begin{equation}
T_{\mu\nu} = \rho u_\mu u_\nu + p h_{\mu\nu} + \pi_{\mu\nu} + 2 q_{(\mu}
u_{\nu)}.  \label{EMT}
\end{equation}

The decomposition is irreducible for any chosen $4$-velocity $u^\mu$. Here $%
\rho$ and $p$ are the energy density and isotropic pressure, and $%
h_{\mu\nu}=g_{\mu\nu}+u_\mu u_\nu$ projects orthogonal to $u^\mu$.

The energy flux obeys the relation $q_\mu=q_{<\mu>}$, and the anisotropic
stress obeys $\pi_{\mu\nu}=\pi_{<\mu\nu>}$, where angular brackets denote
the projected, symmetric and trace-free part:
\begin{equation}
V_{<\mu>} = h_\mu{}^\nu V_\nu, \qquad W_{<\mu\nu>} = \left[
h_{(\mu}{}^\alpha h_{\nu)}{}^\beta - \frac13 h^{\alpha\beta} h_{\mu\nu} %
\right] W_{\alpha\beta}.
\end{equation}

The symmetry properties of $E_{\mu\nu}$ imply that in general we
can decompose it irreducibly with respect to a chosen $4$-velocity
field $u^\mu$ as \cite{Ma01}
\begin{equation}
E_{\mu\nu} = -k^4 \left[ \mathcal{U} \left( u_\mu u_\nu + \frac13
h_{\mu\nu}\right) + \mathcal{P}_{\mu\nu} + 2\mathcal{Q}_{(\mu} u_{\nu)} %
\right],  \label{WT}
\end{equation}
where $k=k_{5}/k_{4}$, $\mathcal{U}$ is a scalar,
$\mathcal{Q}_{\mu }$ a spatial vector and $\mathcal{P}_{\mu \nu }$
a spatial, symmetric and
trace-free tensor. For homogeneous models $\mathcal{Q}_{\mu }=0$ and $%
\mathcal{P}_{\mu \nu }=0$ \cite{CS201}. Hence the only non-zero
contribution from the 5-dimensional Weyl tensor from the bulk is
given by the scalar or ''dark radiation'' term $\mathcal{U}$.

 We consider that in Eq. (\ref{EMT}) the heat transfer is zero,
$q_{\mu }=0$ and we also neglect the effect of the possible
anisotropic stresses (viscosity of the cosmological fluid,
magnetic fields etc.) on the dynamics of the Universe. This allows
to assume that the cosmological fluid on the brane is perfect,
without dissipative effects. Then the matter correction terms are
given by
\begin{equation}
S_{\mu \nu }=\frac{1}{12}\rho ^{2}u_{\mu }u_{\nu }+\frac{1}{12}\rho (\rho
+2p)h_{\mu \nu }.
\end{equation}

The line element of a Bianchi type I space-time, which generalizes
the flat Friedmann-Robertson-Walker metric to the anisotropic
case, is given by
\begin{equation}
ds^{2}=-dt^{2}+a_{1}^{2}\left( t\right) dx^{2}+a_{2}^{2}\left( t\right)
dy^{2}+a_{3}^{2}\left( t\right) dz^{2}.  \label{5}
\end{equation}

We define the following variable:
\begin{equation}  \label{v}
V=\prod_{i=1}^{3}a_{i},
\end{equation}
\begin{equation}  \label{6}
H=\frac{1}{3}\sum_{i=1}^{3}H_{i},
\end{equation}
\begin{equation}  \label{7}
H_{i}=\frac{\dot{a}_{i}}{a_{i}},i=1,2,3,
\end{equation}
\begin{equation}  \label{hi}
\Delta H_{i}=H_{i}-H, i=1,2,3.
\end{equation}

In equations (\ref{v})-(\ref{hi}) $V$ is the volume scale factor, $%
H_{i},i=1,2,3$ are the directional Hubble parameters and $H$ is the mean
Hubble parameter. From Eqs. (\ref{v}) and (\ref{7}) we obtain the important
relation $H=\dot{V}/3V$.

The physical quantities of observational interest in cosmology are the
expansion scalar $\theta $, the mean anisotropy parameter $A$ and the shear
scalar $\sigma ^{2}$, and are defined according to
\begin{equation}
\theta =3H,
\end{equation}
\begin{equation}
A=\frac{1}{3}\sum_{i=1}^{3}\left( \frac{\Delta H_{i}}{H}\right) ^{2},
\end{equation}
\begin{equation}
\sigma ^{2}=\frac{1}{2}\sigma _{ik}\sigma ^{ik}=\frac{1}{2}%
\sum_{i=1}^{3}H_{i}^{2}-3H^{2}=\frac{3}{2}AH^{2}.  \label{10}
\end{equation}

An important observational quantity is the deceleration parameter
\begin{equation}
q=\frac{d}{dt}\left( \frac{1}{H}\right) -1.
\end{equation}

The sign of the deceleration parameter indicates whether the model inflates
or not. The positive sign of $q$ corresponds to ``standard'' decelerating
models whereas the negative sign indicates inflation.

We assume that the isotropic pressure $p$ of the cosmological
fluid on the brane obeys a modified Chaplygin gas equation of
state \cite{De04},
\begin{equation}
p=\gamma \rho -\frac{B}{\rho ^{\alpha }},  \label{chap}
\end{equation}%
where $0\leq \gamma \leq 1$ and $0\leq \alpha \leq 1$. $B$ is a positive
constant.

When $\gamma =1/3$ and the comoving volume of the Universe is
small ($\rho \rightarrow \infty$), this equation of state
corresponds to a radiation dominated era. The main reason for the
use of the modified Chaplygin gas equation of state is that it can
also describe the very high density ($\gamma =1$) and the
radiation dominated phase in the evolution of the Universe. When
the density is small, $\rho \rightarrow 0$, the equation of state
corresponds to a cosmological fluid with negative pressure (the
dark energy). In this limit, the modified Chaplygin equation of
state corresponds to a mixture of cold dark matter and dark energy
\cite{Be02}. At all stages it shows a mixture of dark energy, dark
matter and radiation. For $\rho =\left(B/\gamma \right)^{1/(\alpha
+1)}$ the matter content is pure dust with $p=0$. In the following
we neglect the effect of baryonic (luminous) matter on the
dynamics of the Universe, which, in a rigorous approach, should be
added separately to the equation of state of the cosmological
matter. But since the baryonic matter represents only $4-5\%$ from
the total amount of energy-matter in the Universe \cite{PeRa03},
\cite{Pa03}, its effect on the large scale cosmological dynamics
can be ignored.

The speed of sound $v_s^2=\partial p/\partial \rho $ in the
Chaplygin gas is given by
\begin{equation}
v_s^2=\gamma (1+\alpha )-\frac{\alpha p}{\rho }.
\end{equation}

Using the variables (\ref{v})-(\ref{hi}), the Einstein gravitational field
equations, the Bianchi identity and the evolution equation for the nonlocal
dark radiation take the form, respectively,
\begin{eqnarray}
3\dot{H}+\sum_{i=1}^{3}H_{i}^{2} &=& \Lambda -\frac{k_{4}^{2}}{2}\left[
\left( 3\gamma +1\right) \rho -\frac{3B}{\rho ^{\alpha }}\right] -\frac{%
k_{4}^{2}}{2\lambda }\left[ \left( 3\gamma +2\right) \rho ^{2}-\frac{3B}{%
\rho ^{\alpha -1}}\right] -\frac{6\mathcal{U}}{k_{4}^{2}\lambda },
\label{dH} \\
\frac{1}{V}\frac{d}{dt}\left( VH_{i}\right) &=& \Lambda -\frac{k_{4}^{2}}{2}%
\left[ \left( \gamma -1\right) \rho -\frac{B}{\rho ^{\alpha }}\right] -\frac{%
k_{4}^{2}}{2\lambda }\left[ \gamma \rho ^{2}-\frac{B}{\rho ^{\alpha -1}}%
\right] +\frac{2\mathcal{U}}{k_{4}^{2}\lambda },i=1.2.3,  \label{dVHi} \\
\dot{\rho}+3H\left( p+\rho \right) &=&0,  \label{drho} \\
\dot{\mathcal{U}}+4H\mathcal{U} &=&0,  \label{dU}
\end{eqnarray}

Since we have assumed a homogeneous geometry, the effective
nonlocal energy density, $\mathcal{P} _{\mu \nu }$, and
$\mathcal{Q}_{\mu }$ vanish on the brane, $\mathcal{P} _{\mu \nu
}=0$ and $\mathcal{Q}_{\mu }=0$, respectively.

\section{General solution of the field equations for the Bianchi type I brane%
}

By using the equation of state of the cosmological matter given by Eq. (\ref%
{chap}), the energy conservation equation Eq. (\ref{drho}) can be
integrated exactly, thus giving the time evolution of the energy
density of the modified Chaplygin gas filled Bianchi type I brane
as
\begin{equation}
\rho =\left[ \frac{B}{1+\gamma }+\frac{C}{V^{\left( 1+\gamma \right) \left(
1+\alpha \right) }}\right] ^{\frac{1}{1+\alpha }}.  \label{d}
\end{equation}%
where $C$ is an arbitrary constant of integration.

Integrating Eq. (\ref{dU}) yields
\begin{equation}
\mathcal{U=U}_{0}V^{-\frac{4}{3}},
\end{equation}
where $\mathcal{U}_{0}$ is an arbitrary constant of integration.

By adding Eqs.(\ref{dVHi}) we find
\begin{equation}
\frac{1}{V}\frac{d}{dt}\left( VH\right) =\frac{\ddot{V}}{3V}= \Lambda -\frac{%
k_{4}^{2}}{2}\left[ \left( \gamma -1\right) \rho -\frac{B}{\rho ^{\alpha }}%
\right] -\frac{k_{4}^{2}}{2\lambda }\left[ \gamma \rho ^{2}-\frac{B}{\rho
^{\alpha -1}}\right] +\frac{2\mathcal{U}}{k_{4}^{2}\lambda },  \label{19}
\end{equation}
or, by using the explicit dependence of the energy density on the comoving
volume $V$ and defining a new function $\ddot{V}=F\left( V\right) $, we can
rewrite Eq. (\ref{19}) in the form
\begin{eqnarray}
\ddot{V} &=&F\left( V\right) =3\Lambda V-\frac{3k_{4}^{2}}{2}V\left\{ \left(
\gamma -1\right) \left[ \frac{B}{1+\gamma }+\frac{C}{V^{\left( 1+\gamma
\right) \left( 1+\alpha \right) }}\right] ^{\frac{1}{1+\alpha }}-B\left[
\frac{B}{1+\gamma }+\frac{C}{V^{\left( 1+\gamma \right) \left( 1+\alpha
\right) }}\right] ^{-\frac{\alpha }{1+\alpha }}\right\}  \nonumber  \label{F}
\\
&&-\frac{3k_{4}^{2}}{2\lambda }V\left\{ \gamma \left[ \frac{B}{1+\gamma }+%
\frac{C}{V^{\left( 1+\gamma \right) \left( 1+\alpha \right) }}\right] ^{%
\frac{2}{1+\alpha }}-B\left[ \frac{B}{1+\gamma }+\frac{C}{V^{\left( 1+\gamma
\right) \left( 1+\alpha \right) }}\right] ^{\frac{1-\alpha }{1+\alpha }%
}\right\} +\frac{6\mathcal{U}_{0}}{k_{4}^{2}\lambda }V^{-\frac{1}{3}}.
\end{eqnarray}

The general solution of Eq. (\ref{F}) is given by
\begin{equation}
t-t_{0}=\int \frac{dV}{\sqrt{2\int F\left( V\right) dV+C_{1}}},  \label{time}
\end{equation}
where $C_{1}$ and $t_{0}$ are arbitrary constant of integration.

Subtracting Eq. (\ref{19}) from Eqs. (\ref{dVHi}) we obtain
\begin{equation}
H_{i}=H+\frac{K_{i}}{V},i=1,2,3,  \label{20}
\end{equation}
with $K_{i},i=1,2,3$ constants of integration satisfying the consistency
condition $\sum_{i=1}^{3}K_{i}=0$. From Eq. (\ref{20}), it is easy to show
that
\begin{equation}
\sum_{i=1}^{3}H_{i}^{2}=3H^{2}+\frac{K^{2}}{V^{2}}  \label{21}
\end{equation}
with $\sum_{i=1}^{3}K_{i}^{2}=K^{2}.$

Therefore for a Bianchi type I induced brane geometry in the
presence of the modified Chaplygin gas the general solution of the
gravitational field equation can be expressed in the following
exact parametric form, with $V\geq 0$ taken as parameter:
\begin{equation}
\theta =\frac{\sqrt{2\int F\left( V\right) dV+C_{1}}}{V},  \label{theta}
\end{equation}%
\begin{equation}
a_{i}=a_{i0}V^{\frac{1}{3}}\exp{\left[K_{i}\int
\frac{dV}{V\sqrt{2\int F\left( V\right)
dV+C_{1}}}\right]},i=1,2,3, \label{scale}
\end{equation}%
\begin{equation}
A=\frac{3K^{2}}{2\int F\left( V\right) dV+C_{1}},  \label{A1}
\end{equation}%
\begin{equation}
\sigma ^{2}=\frac{K^{2}}{2V^{2}},  \label{sigma}
\end{equation}%
\begin{equation}
q=2-\frac{3VF\left( V\right) }{2\int F\left( V\right) dV+C_{1}},  \label{q}
\end{equation}%
where $a_{i0},$ $i=1,2,3$, are arbitrary constant of integration.
In order for the general solution of the field equations given by
Eqs. (\ref{theta})-(\ref{q}) be defined for all $V\geq 0$, it is
necessary to chose the constants $\alpha $ and $B$, describing the
thermodynamic properties of the Chaplygin gas and the integration
constant $C$ so that the condition $2\int F\left( V\right)
dV+C_{1}\geq 0$ holds for all $V\geq 0$. Taking into account the
definition of $V$, given by Eq. (\ref{v}), and Eqs. (
\ref{scale}), and since the integration constants $K_{i}$,
$i=1,2,3$ satisfy the condition $\sum_{i=1}^{3}K_{i}=0$, it
follows that the integration constants $a_{i0}$, $i=1,2,3$, have
to satisfy the condition $\prod_{i=1}^{3}a_{i0}=1$.

By introducing two new variables $\tau $ and $v$ by means of the
transformations $\tau =\sqrt{3}k_{4}B^{1/2\left( 1+\alpha \right) }t$ and $%
V=V_{0}v$, with $V_{0}=$constant, by normalizing the values of the arbitrary
integration constants so that $C/BV_{0}^{(1+\gamma )(1+\alpha )}=1$, $4%
\mathcal{U}_{0}V_{0}^{-4/3}/k_{4}^{4}B^{1/(1+\alpha )}\lambda =1$, $%
C_{1}/3V_{0}^{2}k_{4}^{2}B^{1/(1+\alpha )}=1$, denoting
$B^{1/(1+\alpha )}/\lambda =b$, $\lambda _{0}=2\Lambda
/3k_{4}^{2}B^{1/(1+\alpha )}$ and
\begin{equation}
f(v)=\frac{1}{1+\gamma }+\frac{1}{v^{(1+\gamma )(1+\alpha )}},
\end{equation}%
it follows that Eq. (\ref{time}) can be written in the form
\begin{equation}
\tau -\tau _{0}=\int \frac{dv}{\sqrt{\int \left\{ \lambda _{0}v-v\left[
\left( \gamma -1\right) f^{\frac{1}{1+\alpha }}-f^{-\frac{\alpha }{1+\alpha }%
}\right] -bv\left[ \gamma f^{\frac{2}{1+\alpha }}-f^{\frac{1-\alpha }{%
1+\alpha }}\right] +v^{-\frac{1}{3}}\right\} dv+1}},
\end{equation}
where $\tau _0=\sqrt{3}k_{4}B^{1/2\left( 1+\alpha \right) }t_0$.

With the use of Eqs. (\ref{d}), (\ref{20}) and (\ref{theta}), from Eq. (\ref%
{dH}) it follows that the arbitrary integration constants $K_{i}$, $i=1,2,3$
and $C_{1}$ must satisfy the consistency condition
\begin{equation}
K^{2}=\frac{2C_{1}}{3},  \label{K1}
\end{equation}

Scalar fields are supposed to play a fundamental role in the
evolution of the early universe. The Chaplygin gas model can be
also described from a field theoretical point of view by
introducing a scalar field $\phi $ and a self interacting
potential $U(\phi )$, with the Lagrangian \cite{Ka01},
\cite{Be02}, \cite{Bi02}, \cite{Ber04}, \cite{De04}
\begin{equation}
L_{\phi }=\frac{1}{2}\dot{\phi}^{2}-U(\phi ).
\end{equation}

The energy density and the pressure associated to the scalar field
$\phi $ are given by
\begin{equation}
\rho _{\phi }=\frac{\dot{\phi}^{2}}{2}+U\left( \phi \right) =\rho ,
\label{rho}
\end{equation}%
and
\begin{equation}
p_{\phi }=\frac{\dot{\phi}^{2}}{2}-U\left( \phi \right) =\gamma \rho -\frac{B%
}{\rho ^{\alpha }},  \label{p}
\end{equation}%
respectively.

In view of Eqs.(\ref{rho}), (\ref{p}) and (\ref{d}) we have
\begin{equation}
\dot{\phi}^{2}=\left( 1+\gamma \right) \left[ \frac{B}{1+\gamma }+\frac{C}{%
V^{\left( 1+\gamma \right) \left( 1+\alpha \right) }}\right] ^{\frac{1}{%
1+\alpha }}-B\left[ \frac{B}{1+\gamma }+\frac{C}{V^{\left( 1+\gamma \right)
\left( 1+\alpha \right) }}\right] ^{-\frac{\alpha }{1+\alpha }},
\label{phi1}
\end{equation}
and
\begin{equation}
U\left( \phi \right) =\frac{\left( 1-\gamma \right) }{2}\left[ \frac{B}{%
1+\gamma }+\frac{C}{V^{\left( 1+\gamma \right) \left( 1+\alpha \right) }}%
\right] ^{\frac{1}{1+\alpha }}+\frac{B}{2}\left[ \frac{B}{1+\gamma }+\frac{C%
}{V^{\left( 1+\gamma \right) \left( 1+\alpha \right) }}\right] ^{-\frac{%
\alpha }{1+\alpha }},  \label{U1}
\end{equation}
respectively.

By integrating Eq. (\ref{phi1}), we obtain
\begin{equation}
\phi -\phi _{0}=\int \sqrt{\frac{\left( 1+\gamma \right) \left[ \frac{B}{%
1+\gamma }+\frac{C}{V^{\left( 1+\gamma \right) \left( 1+\alpha \right) }}%
\right] ^{\frac{1}{1+\alpha }}-B\left[ \frac{B}{1+\gamma }+\frac{C}{%
V^{\left( 1+\gamma \right) \left( 1+\alpha \right) }}\right] ^{-\frac{\alpha
}{1+\alpha }}}{2\int F\left( V\right) dV+C_{1}}}dV,  \label{phi2}
\end{equation}
where $\phi _{0}$ is an arbitrary constant of integration.

\section{Discussions and final remarks}

In the present paper we have considered Chaplygin gas filled
Bianchi type I brane worlds in the framework of the
Randall-Sundrum scenario \cite{RS99a}. The Randall-Sundrum
mechanism was originally motivated as a possible mechanism for
evading Kaluza-Klein compactification by localizing gravity in the
presence of an uncompactified extra dimension. In this model, the
5-dimensional bulk space-time is assumed to be vacuum except for
the presence of a cosmological constant. Matter fields on the
brane are regarded as responsible for the dynamics of the brane.
The quadratic terms in the energy-momentum tensor, due to the form
of the Gauss-Codazzi equations, may be important in the very early
stages of the Universe, leading to major changes in the dynamics
of the Universe. The general solution of the gravitational field
equations on the brane can be obtained for both discussed brane
geometries in an exact parametric form, with the comoving volume
element $V$ taken as parameter.

By obtaining the general solution of the field equations we have assumed the
presence of a non-zero effective cosmological constant $\Lambda $ on the
brane. However, the most attractive feature of the Chaplygin gas is that it
could explain the main observational properties of the Universe without
appealing to an effective cosmological constant. To see if this is indeed
the case in the anisotropic brane world models, in the following we will
assume that on the brane $\Lambda =0$. This constraint imposes the following
relation between the vacuum energies in the bulk and on the brane:
\begin{equation}
\Lambda _5=-\frac{k_5^2}{6}\lambda ^2
\end{equation}

Therefore for $\lambda _0=0$ and for a fixed $\gamma $ and $\alpha $, the
dynamics of the anisotropic brane Universes is controlled by the parameter $%
b=B^{1/(1+\alpha )}/\lambda $, the ratio of the Chaplygin gas equation of
state parameter and the vacuum energy on the brane.

From the equation of state of the Chaplygin gas with $\gamma =0$
it follows that for the critical values $p_{c}$ and $\rho _{c}$ of
the pressure and density the parameter $w_{c}=p_{c}/\rho _{c}$ is
given by $w_{c}=-B/\rho
_{c}^{\alpha +1}$. Evaluating this relation at the present time gives $%
B=-w_{c0}\rho _{c0}^{\alpha +1}$. The Chaplygin gas behaves like a
cosmological constant for $w_{c0}=-1$, which fixes the constant $B$ as $%
B=\rho _{c0}^{\alpha +1}=\left( 3H_{0}^{2}/8\pi G\right) ^{\alpha
+1}$, where $H_{0}=3.24\times 10^{-18}h$ s$^{-1}$, $0.5\leq h\leq
1$, is the
Hubble constant \cite{PeRa03}, \cite{Pa03}. The definition of the dimensionless time parameter $\tau =%
\sqrt{3}k_{4}B^{1/2\left( \alpha +1\right) }t$ then gives $\tau
=3H_{0}t$, or $t\approx 10^{17}h^{-1}\times \tau $ s$^{-1}$.

Generally, the obtained solution of the gravitational field
equations describes an expanding brane Universe. The variation of
the expansion scalar $\theta $ as a function of the dimensionless
time parameter $\tau $ is represented in Fig. 1.

\vspace{0.2in}
\begin{figure}[h]
\includegraphics{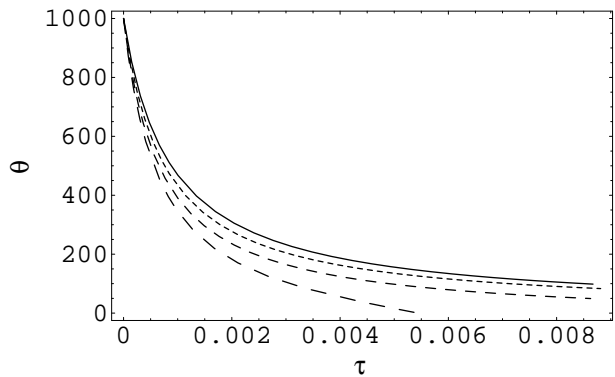}
\caption{Expansion scalar $\theta $ of the Chaplygin gas filled
Bianchi type I brane universe as a function of the dimensionless
time parameter $\tau $ for $\gamma =1/3$, $\alpha =0.5$ and
$b=0.01$ (solid curve), $b=0.02$ (dotted curve), $b=0.03$ (dashed
curve) and $b=0.04$ (long dashed curve). We have assumed a
vanishing effective cosmological constant on the brane,
$\protect\lambda _0=0$. The relation between $\tau $ and the
cosmic time $t$ is $t\approx 10^{17}h^{-1}\times \tau $ s$^{-1}$,
$0.5\leq h\leq 1$.} \label{FIG1}
\end{figure}

For a Chaplygin type cosmological fluid confined on the brane, the early
cosmological evolution in anisotropic Bianchi type I geometry is
fundamentally changed by the inclusion of the terms proportional to the
square of the energy density and energy density-pressure products. The time
variation of the mean anisotropy parameter of the Bianchi type I space-time
is represented, for $\gamma =1/3$, $\alpha =0.5$ and for different values of
the parameter $b$, in Fig. 2. At high densities the brane Universe starts
its evolution from an isotropic state, with $A(t_0)=0$. The anisotropy
increases and reaches a maximum value after a finite time interval $t_{\max}$%
, and for $t>t_{\max}$, the mean anisotropy is a monotonically decreasing
function, tending to zero in the large time limit. This behavior is in sharp
contrast to the usual general relativistic evolution (in this case the
quadratic contribution to the energy-momentum tensor vanishes), in which the
Universe is born in a state of maximum anisotropy.

\vspace{0.2in}
\begin{figure}[h]
\includegraphics{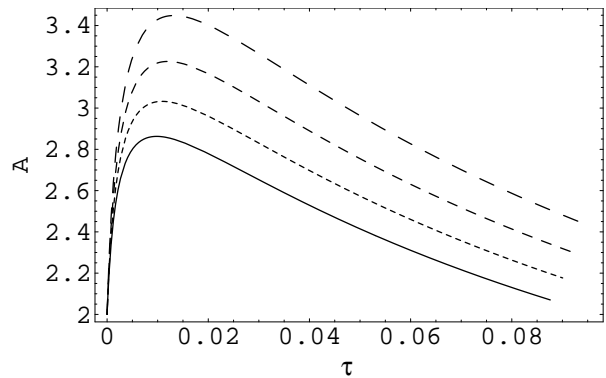}
\caption{Anisotropy parameter $A$ of the Chaplygin gas filled Bianchi type I
brane universe as a function of the dimensionless time parameter $\protect%
\tau $ for $\protect\gamma =1/3$, $\protect\alpha =0.5$ and
$b=0.01$ (solid curve), $b=0.011$ (dotted curve), $b=0.012$
(dashed curve) and $b=0.013$ (long dashed curve). We have assumed
a vanishing effective cosmological constant on the brane, $\lambda
_0=0$. The relation between $\tau $ and the cosmic time $t$ is
$t\approx 10^{17}h^{-1}\times \tau $ s$^{-1}$, $0.5\leq h\leq 1$.}
\label{FIG2}
\end{figure}

The time variation of the shear parameter $\sigma ^2$ is
represented, as a function of the dimensionless time $\tau $ and
for different values of $b$, in Fig. 3.

\vspace{0.2in}
\begin{figure}[h]
\includegraphics{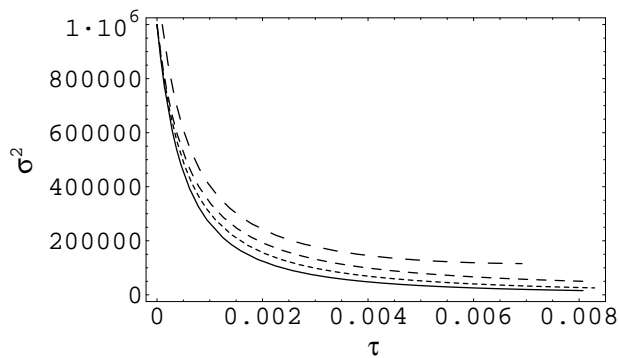}
\caption{Shear scalar $\sigma ^2$ of the Chaplygin gas filled
Bianchi type I brane universe as a function of the dimensionless
time parameter $\tau $ for $\gamma =1/3$, $\alpha =0.5$ and
$b=0.01$ (solid curve), $b=0.02$ (dotted curve), $b=0.03$ (dashed
curve) and $b=0.04$ (long dashed curve). We have assumed a
vanishing effective cosmological constant on the brane, $\lambda
_0=0$. The relation between $\tau $ and the cosmic time $t$ is
$t\approx 10^{17}h^{-1}\times \tau $ s$^{-1}$, $0.5\leq h\leq 1$.}
\label{FIG3}
\end{figure}

The shear is a monotonically decreasing function of the
cosmological time, and in the limit of large $t$, $t\rightarrow
\infty$, corresponding to the isotropic limit, $\sigma ^2$ tends
to zero. In the limit of small times, since $\sigma ^2\propto
V^{-2}$, the shear has a singular behavior, tending to infinity,
$\lim_{t\rightarrow 0}\sigma ^2\rightarrow \infty$.

In the initial stage the evolution of the Chaplygin gas filled
Bianchi type I brane universe is non-inflationary, but in the
large time limit the brane Universe ends in an accelerating stage.
In Fig. 4 we
present the dynamics of the deceleration parameter for different values of $%
b $ and for $\gamma =1/3$ and $\alpha =0.5$.

\vspace{0.2in}
\begin{figure}[h]
\includegraphics{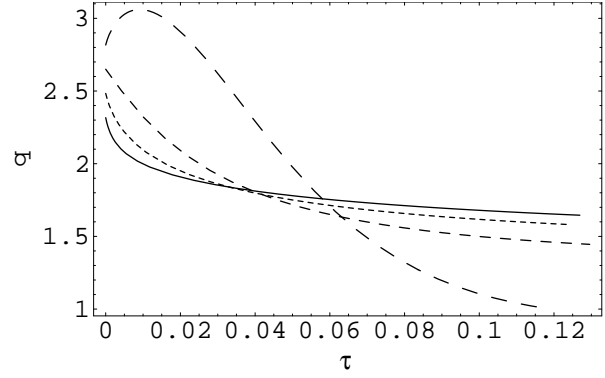}
\caption{Deceleration parameter $q$ of the Chaplygin gas filled Bianchi type
I brane universe as a function of the dimensionless time parameter $\protect%
\tau $ for $\gamma =1/3$, $\alpha =0.5$ and $b=0.01$ (solid
curve), $b=0.015$ (dotted curve), $b=0.02$ (dashed curve) and
$b=0.025$ (long dashed curve). We have assumed a vanishing
effective cosmological constant on the brane, $\protect\lambda
_0=0$. The relation between $\tau $ and the cosmic time $t$ is
$t\approx 10^{17}h^{-1}\times \tau $ s$^{-1}$, $0.5\leq h\leq 1$.}
\label{FIG4}
\end{figure}

The time variation of the scalar field associated with the
Chaplygin gas is represented for fixed $\alpha $ and $\gamma $ and
different values of the parameter $b$ in Fig. 5. The scalar field
is a monotonically decreasing function of time.

\vspace{0.2in}
\begin{figure}[h]
\includegraphics{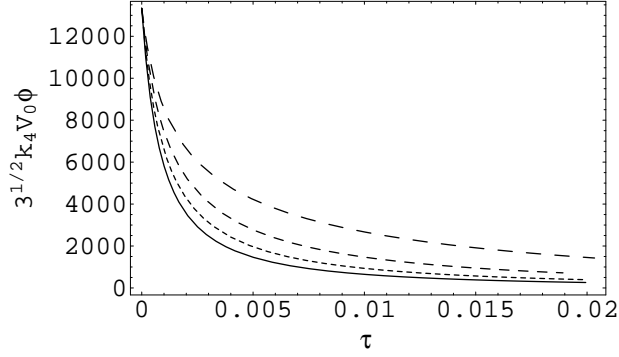}
\caption{Chaplygin gas associated scalar field $\phi $ evolution
on the Bianchi type I brane universe as a function of the
dimensionless time parameter $\tau $ for $\protect\gamma =1/3$,
$\alpha =0.5$ and $b=0.005$ (solid curve), $b=0.01$ (dotted
curve), $b=0.015$ (dashed curve) and $b=0.02$ (long dashed curve).
We have assumed a vanishing effective cosmological constant on the
brane, $\protect\lambda _0=0$. The relation between $\tau $ and
the cosmic time $t$ is $t\approx 10^{17}h^{-1}\times \tau $
s$^{-1}$, $0.5\leq h\leq 1$.} \label{FIG5}
\end{figure}

The dependence of the potential $U(\phi )$ on the scalar field
$\phi $ is represented in Fig. 6.

\vspace{0.2in}
\begin{figure}[h]
\includegraphics{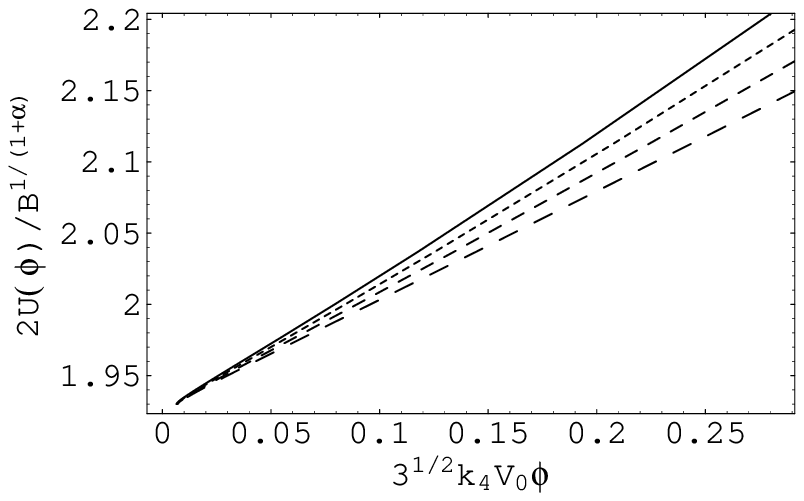}
\caption{The potential $U(\protect\phi )$ of the Chaplygin gas associated
scalar field as a function of the scalar field $\protect\phi $ for $\protect%
\gamma =1/3$, $\protect\alpha =0.5$ and $b=0.005$ (solid curve), $b=0.01$
(dotted curve), $b=0.015$ (dashed curve) and $b=0.02$ (long dashed curve).
We have assumed a vanishing effective cosmological constant on the brane, $%
\protect\lambda _0=0$.} \label{FIG6}
\end{figure}

In the limit of high densities, the equation of state of the Chaplygin gas
reduces to the linear barotropic equation of state of the relativistic
matter, $p=\gamma \rho $. In this case the density dependence on the
comoving volume is given by $\rho =\rho _{0}/V^{\gamma +1}$, $\rho _{0}=$%
constant, and the equation describing the time variation of the comoving
volume reduces to
\begin{equation}
t-t_{0}=\int \frac{dV}{\sqrt{3k_{4}^{2}\rho _{0}V^{1-\gamma }+\frac{%
3k_{4}^{2}}{2\lambda }\rho _{0}^{2}V^{-2\gamma }+\frac{18\mathcal{U}_{0}}{%
k_{4}^{2}\lambda }V^{2/3}+C_{1}}}.  \label{low}
\end{equation}

In the limit of very high matter densities, which can be described by the
strong equation of state for the hot nucleonic gas, it is assumed that $%
p=\rho $. It is believed that matter actually behaves in this
manner at densities above about ten times the nuclear density,
that is, at densities greater than $10^{17}$ g/cm$^{3}$ and at
temperatures $T=\left( \rho c^2
/\sigma _{rad} \right) ^{1/4}>10^{13}$ K, where $\sigma _{rad}$ is the radiation constant %
\cite{ZN71}. Therefore in this case $\gamma =1$. In the limit of small $V$,
the term $V^{-2}$ will dominate in the integral in Eq. (\ref{low}). By
maintaining only this term, the integration can be done easily and the time
dependence of the relevant geometrical and physical parameters can be
obtained as
\begin{equation}
V=\left( \frac{6k_{4}^{2}\rho _{0}^{2}}{\lambda }\right) ^{1/4}t^{1/2},
\end{equation}%
\begin{equation}
\theta =\frac{1}{2t},
\end{equation}
\begin{equation}
a_{i}=a_{i0}t^{1/6}\exp \left[ 2K_{i}\left( \frac{%
6k_{4}^{2}\rho _{0}^{2}}{\lambda }\right) ^{-1/4}t^{1/2}\right]
,i=1,2,3,
\end{equation}
\begin{equation}
A=\frac{12K^{2}}{\left( \frac{6k_{4}^{2}\rho _{0}^{2}}{\lambda }\right)
^{1/2}}t,
\end{equation}%
\begin{equation}
\sigma ^{2}=\frac{K^{2}}{2\left( \frac{6k_{4}^{2}\rho _{0}^{2}}{\lambda }%
\right) ^{1/2}}\frac{1}{t},
\end{equation}%
\begin{equation}
q=5,
\end{equation}%
where we have assumed the initial conditions $t_{0}=0$ and $V(0)=0$.

In the opposite limit of small densities, the equation of state of the
Chaplygin gas becomes $p=-B/\rho ^{\alpha }$, corresponding to the limit $%
\gamma \rightarrow 0$. In this case the equation describing the time
dependence of the comoving volume $V$ is given by
\begin{equation}
\ddot{V}=\frac{3k_{4}^{2}}{2}V\left\{ \left[ B+\frac{C}{V^{1+\alpha }}\right]
^{\frac{1}{1+\alpha }}+B\left[ B+\frac{C}{V^{1+\alpha }}\right] ^{-\frac{%
\alpha }{1+\alpha }}\right\} +\frac{3k_{4}^{2}B}{2\lambda }V\left[ B+\frac{C%
}{V^{1+\alpha }}\right] ^{\frac{1-\alpha }{1+\alpha }}+\frac{6\mathcal{U}_{0}%
}{k_{4}^{2}\lambda }V^{-\frac{1}{3}}.  \label{high}
\end{equation}

In the limit of large $V$ the condition $B>>C/V^{1+\alpha }$ is
always fulfilled, and the general solution of Eq. ( \ref{high})
can be expressed as
\begin{equation}
t-t_{0}=\int \frac{dV}{\sqrt{\Lambda _{c}^{2}V^{2}+\frac{18\mathcal{U}_{0}}{%
k_{4}^{2}\lambda }V^{2/3}+C_{1}}},  \label{high1}
\end{equation}
where we have denoted
\begin{equation}
\Lambda _{c}=\sqrt{3k_{4}^{2}B^{1/(1+\alpha )}\left( 1+\frac{B^{1/(1+\alpha
)}}{2\lambda }\right) }.
\end{equation}

In the same limit $V\rightarrow \infty $ we also have $\Lambda _{c}V^{2}>>18%
\mathcal{U}_{0}V^{2/3}/k_{4}^{2}\lambda +C_{1}$ and therefore Eq. (\ref%
{high1}) can be immediately integrated to give
\begin{equation}
V=V_{1}\exp \left[ \Lambda _{c}\left( t-t_{0}\right) \right] .
\end{equation}
where $V_{1}$ is an arbitrary constant of integration.

The time evolution of the physical and geometrical quantities in
the limit of the large comoving volume is given by
\begin{equation}
\theta =\Lambda _{c},
\end{equation}%
\begin{equation}
a_{i}=a_{i0}\exp \left[
\frac{\Lambda _{c}}{3}\left( t-t_{0}\right) \right] \exp \left\{ -\frac{K_{i}%
}{\Lambda _{c}V_{1}}\exp \left[ -\Lambda _{c}\left( t-t_{0}\right)
\right] \right\} ,i=1,2,3,
\end{equation}
\begin{equation}
A=\frac{3K^{2}}{%
\Lambda _{c}^{2}V_{1}^{2}}\exp \left[ -2\Lambda _{c}\left(
t-t_{0}\right)\right]
\end{equation}
\begin{equation}
\sigma ^{2}=\frac{K^{2}}{2V_{1}^{2}}\exp \left[ -2\Lambda
_{c}\left( t-t_{0}\right) \right],
\end{equation}
\begin{equation}
q=-1.
\end{equation}

In the limit of large times the mean anisotropy parameter tends to zero, $%
A\rightarrow 0$, and the Chaplygin gas filled Bianchi type I
Universe ends in an isotropic state, corresponding to a de Sitter
type Universe. This shows that the Chaplygin gas model offers a
real possibility for replacing the effective cosmological constant
on the brane.


\begin{references}

\bibitem{Pe99} S. Perlmutter et al., Astrophys. J. {\bf 517}, 565
(1998); A. G. Riess et al., Astron. J. {\bf 116}, 109 (1998).

\bibitem{Ber00} P. de Bernardis et al., Nature {\bf 404}, 995
(2000); S. Hanany et al., Astrophys. J. {\bf 545}, L5 (2000).

\bibitem{PeRa03} P. J. E. Peebles and B. Ratra, Rev. Mod. Phys.
{\bf 75}, 559 (2003).

\bibitem{Pa03} T. Padmanabhan, Phys. Repts. {\bf 380}, 235 (2003).

\bibitem{Paetal} T. Matos and A. Urena-Lopez, Class. Quantum Grav.
{\bf 17}, L75 (2000); C. Wetterich, Phys. Rev. D. {\bf 65},
123512; T. Padmanabhan and T. R. Choudhury, Phys. Rev. D {\bf 66},
081301 (2002).

\bibitem{Ka01} A. Kamenshchik, U. Moschella and V. Pasquier, Phys.
Lett. B {\bf 511}, 265 (2001).

\bibitem{Be02}M. C. Bento, O. Bertolami and A. Sen, Phys. Rev. D {\bf 66},
043507 (2002).

\bibitem{Ka00}A. Kamenshchik, U. Moschella and V. Pasquier, Phys.
Lett. B {\bf 487}, 7 (2000).

\bibitem{Ka98} S. K. Kama, Phys. Lett. B {\bf 424}, 39 (1998).

\bibitem{Bi02} N. Bilic, G. B. Tupper and R. D. Viollier, Phys.
Lett. B {\bf 535}, 17 (2002).

\bibitem{No05}  M. Novello, M. Makler, L. S. Werneck and C. A.
Romero, Phys. Rev. D {\bf 71}, 043515 (2005).

\bibitem{Chco}  D. Carturan and F. Finelli, Phys. Rev. D {\bf 68}, 103501 (2003);
R. Bean and O. Dore, Phys. Rev. D {\bf 68}, 023515 (2003); L. M.
G. Beca, P. P. Avelino, J. P. M. de Carvalho and C. J. A. P.
Martins, Phys. Rev. D {\bf 67}, 101301  (2003); M. C. Bento, O.
Bertolami and A. A. Sen, Phys. Rev. D {\bf 70} 083519,  (2004); R.
R. R. Reis, I. Waga, M. O. Calvao and S. E.
 Joras, Phys. Rev. D {\bf 68} 061302 (2003); P. P. Avelino, L. M. G. Beca, J. P. M. de Carvalho and C. J. A. P.
 Martins, JCAP {\bf 0309} 002 (2003);
G. M. Kremer, Phys. Rev. D {\bf 68} 123507 (2003); T. Multamaki,
M. Manera and E. Gaztanaga, Phys. Rev. D {\bf 69}, 023004 (2004);
 M. Szydlowski and W. Czaja, Phys. Rev. D {\bf 69}
 023506 (2004); P. P. Avelino, L. M. G. Beca, J. P. M. de Carvalho, C. J. A. P. Martins and E.J.
 Copeland, Phys. Rev. D {\bf 69} 041301 (2004).

\bibitem{Ber04} O. Bertolami, A. A. Sen, S. Sen and P. T. Silva, Mon. Not. Roy.
Astron. Soc. {\bf 353}, 329 (2004)

 \bibitem{Fa02}  M. Makler, S. Quinet de Oliveira and I. Waga, Phys. Lett. B {\bf 555},
 1 (2003); J. V. Cunha, J. S. Alcaniz, J. A. S. Lima, Phys. Rev. D {\bf 69},
 083501 (2004); M. C. Bento, O. Bertolami, N. M. C. Santos and A. A. Sen, Phys. Rev. D {\bf 71}, 063501 (2005).

\bibitem{Be03} M. C. Bento, O. Bertolami and A. A. Sen, Phys. Lett. B {\bf 575},
172 (2003);  M. C. Bento, O. Bertolami and A. A. Sen, Phys. Rev. D
{\bf 67}, 063003 (2003);  L. Amendola, F. Finelli, C. Burigana, D.
Carturan, JCAP {\bf 0307}, 005 (2003).

\bibitem{De03} A. Dev, D. Jain and J. S. Alcaniz, Phys. Rev. D
{\bf 67}, 023515 (2003); A. Dev, D. Jain and J. S. Alcaniz,
Astron. Astrophys. {\bf 417}, 847 (2004); P. T. Silva and O.
Bertolami, Astrophys. J. {\bf 599}, 829 (2003).

\bibitem{Al03} J. S. Alcaniz, D. Jain and A. Dev, Phys. Rev. D
{\bf 67}, 043514 (2003).

\bibitem{Cu04} J. V. Cunha, J. A. S. Lima and J. S. Alcaniz, Phys.
Rev. D {\bf 69}, 083501 (2004); Z.-H. Zhu, Astron. Astrophys. {\bf
423}, 421 (2004).

\bibitem{Ei} L. P. Eisenhart, Riemannian geometry, Princeton,
Princeton University Press (1949).

\bibitem{Ak83} K. Akama, Pregeometry, in Lect. Notes Phys. {\bf
176}, 267 (1983).

\bibitem{RuSh83} V. A. Rubakov and M. E. Shaposhnikov, Phys. Lett.
B{\bf 125}, 136 (1983).

\bibitem{BaKu99} I. Bandos and W. Kummer, Int. J. Mod. Phys. A {\bf
14}, 4881 (1999).

\bibitem{PaTa00} M. Pavsic and V. Tapia, gr-qc/0010045 (2000).

\bibitem{RS99a}  L. Randall and R. Sundrum, Phys. Rev. Lett. {\bf 83},
3370 (1999); L. Randall and R. Sundrum, Phys. Rev. Lett {\bf 83},
4690 (1999).

\bibitem{Ma01}  R. Maartens, Living Rev. Rel. {\bf 7}, 7 (2004).

\bibitem{SMS00}  T. Shiromizu, K. Maeda and M. Sasaki, {\it Phys. Rev. {\bf %
D62}}, 024012 (2000); M. Sasaki, T. Shiromizu and K. Maeda, {\it Phys. Rev. {\bf %
D62}}, 024008 (2000).

\bibitem{all}  K. Maeda and D. Wands, Phys. Rev. D {\bf 62}, 124009
(2000); R. Maartens, Phys. Rev. D {\bf 62}, 084023 (2000); A.
Campos and C. F. Sopuerta, Phys. Rev. D{\bf 64}, 104011 (2001);
 D. Langlois, Phys. Rev. Lett. {\bf 86}, 2212 (2001); C.-M.
Chen, T. Harko and M. K. Mak, Phys. Rev. D {\bf 64}, 124017
(2001); L. Anchordoqui, J. Edelstein, C. Nunez, S. P. Bergliaffa,
M. Schvellinger, M. Trobo and F. Zyserman, Phys. Rev. D {\bf 64},
084027 (2001);  J. D. Barrow and R. Maartens, Phys. Lett. B {\bf
532}, 153 (2002); H. Kudoh and T. Tanaka, Phys. Rev. D {\bf 65},
104034 (2002); H. A. Bridgman, K. A. Malik and D. Wands, Phys.
Rev. D {\bf 65},
043502 (2002); C.-M. Chen, T. Harko, W. F. Kao and M. K. Mak, Nucl. Phys. B {\bf %
636}, 159 (2002); M. Szydlowski, M. P. Dabrowski and A. Krawiec,
Phys. Rev. D {\bf 66}, 064003 (2002); T. Harko and M. K. Mak ,
Class. Quantum Grav. {\bf 20}, 407 (2003), C.-M. Chen, T. Harko,
W. F. Kao and M. K. Mak, JCAP {\bf 0311}, 005 (2003);   Iver
Brevik, Kazuo Ghoroku, Masanobu Yahiro, Phys. Rev. D {\bf 70}
064012, (2004).

\bibitem{all1}  N. Dadhich and S. G. Ghosh, Phys. Lett. {\bf B518}, 1
(2001); M. G. Santos, F. Vernizzi and P. G. Ferreira, Phys. Rev. D {\bf %
64}, 063506 (2001); M. Bruni, C. Germani and R. Maartens, Phys.
Rev. Lett. {\bf 87}, 231302 (2001); H.-C. Kim, S.-H. Moon and J.
H. Yee, JHEP {\bf 0202}, 046 (2002); M Govender and N. Dadhich,
 Phys. Lett. B {\bf 538}, 233 (2002); T. Wiseman, Class.
Quant. Grav. {\bf 19}, 3083 (2002); R. Neves and C. Vaz, Phys.
Rev. D {\bf 66}, 124002 (2002); L. A. Anchordoqui, H. Goldberg and
A. D. Shapere, Phys. Rev. D {\bf 66}, 024033 (2002); H. Kudoh, T.
Tanaka and T. Nakamura, Phys. Rev. D {\bf 68}, 024035 (2003); T.
Harko and M. K. Mak,  Phys. Rev. D {\bf 69}, 064020 (2004).

\bibitem{MaHa04} M. K. Mak and T. Harko, Phys. Rev. D {\bf 70},
024010 (2004).

\bibitem{MSS00}R. Maartens, V. Sahni and T. D. Saini,Phys. Rev. D {\bf 63},
063509 (2001); A. V. Toporensky, Class. Quant. Grav. {\bf 18},
2311 (2001);   A. V. Frolov, Phys. Lett. B {\bf 514}, 213 (2001);
M. G. Santos, F. Vernizzi and P. G. Ferreira, Phys. Rev. D {\bf
64}, 063506 (2001);  T. Harko and M. K. Mak, Class. Quant. Grav.
{\bf 20}, 407 (2003).

\bibitem{ChHaMa01a}C.-M. Chen, T. Harko and M. K. Mak, Phys. Rev. D {\bf 64}, 044013
(2001).

\bibitem{CS201} A. Campos and C. F. Sopuerta, Phys. Rev. D {\bf 63},
104012.

\bibitem{Co01a}  A. Coley, Phys. Rev. D {\bf 66} 023512 (2002).

\bibitem{Co01b}  A. Coley,  Class. Quantum Grav. {\bf 19},
L45 (2002).

\bibitem{HaMa04a}T. Harko and M. K. Mak, Class. Quantum. Grav. {\bf
21}, 1489 (2004).

\bibitem{MW00} K. Maeda and D. Wands, Phys. Rev. D {\bf 62}, 124009
(2000).

\bibitem{De04} U. Debnath, A. Banerjee and S. Chakraborty, Class.
Quant. Grav. {\bf 21},  5609 (2004).

\bibitem{ZN71} Ya. B.  Zeldovich and I. D. Novikov,
 Relativistic Astrophysics, Univ. of Chicago Press, Chicago,
Ill (1971).

\end{references}
\end{document}